\documentclass[11pt]{amsart}
\usepackage[colorlinks = true,
            linkcolor = red,
            urlcolor  = magenta,
            citecolor = black,
            anchorcolor = blue]{hyperref}
\usepackage{color}
\usepackage[alphabetic]{amsrefs}
\usepackage{tikz}
\usetikzlibrary{arrows}
\usepackage[all]{xy}
\usepackage{graphicx}
\usepackage{wrapfig}
\usepackage{youngtab}
\usepackage{geometry}                
\usepackage{graphicx}
\usepackage{amssymb}
\usepackage{epstopdf}
\usepackage{mathrsfs}
\usepackage[mathcal]{eucal}

\makeatletter \@addtoreset{equation}{section} \makeatother

\renewcommand{\eprint}[1]{\href{https://arxiv.org/abs/#1}{#1}}

\BibSpec{article}{%
    +{}  {\PrintAuthors}                {author}
    +{,} { \textit}                     {title}
    +{.} { }                            {part}
    +{:} { \textit}                     {subtitle}
    +{,} { \PrintContributions}         {contribution}
    +{.} { \PrintPartials}              {partial}
    +{,} { }                            {journal}
    +{}  { \textbf}                     {volume}
    +{}  { \PrintDatePV}                {date}
    +{,} { \issuetext}                  {number}
    +{,} { \eprintpages}                {pages}
    +{,} { }                            {status}
    +{,} { \url}                        {url}    
    +{,} { \DOI}                   {doi}
    +{,} { \eprint}        {eprint}
    +{}  { \parenthesize}               {language}
    +{}  { \PrintTranslation}           {translation}
    +{;} { \PrintReprint}               {reprint}
    +{.} { }                            {note}
    +{.} {}                             {transition}
    +{}  {\SentenceSpace \PrintReviews} {review}
}

\newcommand{\CZ}{\mathcal{Z}}

\makeatletter
\newcommand{\ellSN}{\mathop{\operator@font sn}\nolimits}
\newcommand{\ellCN}{\mathop{\operator@font cn}\nolimits}
\newcommand{\ellDN}{\mathop{\operator@font dn}\nolimits}
\newcommand{\ellAM}{\mathop{\operator@font am}\nolimits}
\newcommand{\ellK}{\mathop{\smash{\operator@font K}\vphantom{a}}\nolimits}
\newcommand{\ellE}{\mathop{\smash{\operator@font E}\vphantom{a}}\nolimits}
\makeatother

\ifx\genfrac\sdflkaj

\else

\fi

\newcommand{\beq}{\begin{equation}}
\newcommand{\eeq}{\end{equation}}

\makeatletter
\def\mr@ignsp#1 {\ifx\:#1\@empty\else #1\expandafter\mr@ignsp\fi}%
\newcommand{\multiref}[1]{\begingroup
\xdef\mr@no@sparg{\expandafter\mr@ignsp#1 \: }%
\def\mr@comma{}%
\@for\mr@refs:=\mr@no@sparg\do{\mr@comma\def\mr@comma{,}\ref{\mr@refs}}%
\endgroup}
\makeatother

\newcommand{\hypref}[2]{\ifx\href\asklfhas #2\else\href{#1}{#2}\fi}

\newcommand{\secref}[1]{Sec.~\multiref{#1}}

\newcommand{\figref}[1]{Fig.~\multiref{#1}}
\renewcommand{\eqref}[1]{(\multiref{#1})}

\def\[{\begin{equation}}
\def\]{\end{equation}}
\def\<{\begin{eqnarray}}
\def\>{\end{eqnarray}}

\ifx\href\asklfhas\newcommand{\href}[2]{#2}\fi

\title{On Quiver W-algebras and Defects from Gauge Origami}

\author{Peter Koroteev}
\address{\newline Peter Koroteev
\newline
Department of Mathematics,\newline
University of California Berkeley,\newline
970 Evans Hall \#3840,\newline
University of California,\newline
Berkeley, CA 94720-3840,\newline
The United States of America}

\begin{document}
\maketitle

\begin{abstract}
In this note, using Nekrasov's gauge origami framework, we study two different versions of the
the BPS/CFT correspondence -- first, the standard AGT duality and, second, the quiver W algebra construction which has been developed recently by Kimura and Pestun. The gauge origami enables us to work with both dualities simultaneously and find exact matchings between the parameters.
In our main example of an $A$-type quiver gauge theory, we show that the corresponding quiver qW-algebra and its representations are closely related to a large-$n$ limit of spherical $\mathfrak{gl}_n$ double affine Hecke algebra whose modules are described by instanton partition functions of a defect quiver theory. 
\end{abstract}


\section{Introduction}\label{Sec:Intro}
The BPS/CFT correspondence \cite{Nekrasov:2015wsu} is an intriguing duality which relates counting of BPS states in gauge theories with extended supersymmetry in various dimensions with conformal field theories and their symmetries, i.e. (q)vertex operator algebras (VOA)s. The most well-studied example is the Alday-Gaiotto-Tachikawa (AGT) correspondence \cite{Alday:2009aq} which conjectures equalities between Liouville (Toda) conformal blocks and Nekrasov partition functions of the dual 4d $\mathcal{N}=2$ gauge theories with eight supercharges. As an example, the instanton partition function of pure $SU(n)$ super Yang-Mills theory is equal to the conformal block of $n$-Toda CFT which has W$_n$ algebra symmetry, therefore the symmetry algebra is directly related to the number of colors of the gauge group.

A different version of the BPS/CFT correspondence can be found in papers by Kimura and Pestun \cite{Kimura:2015rgi,Kimura:2016dys} which can naturally be formulated in five dimensions for quiver gauge theories whose quivers have shapes of Dynkin diagrams of root system $\Gamma$.  According to their construction, which will be reviewed in \secref{Sec:QuiverWRev}, the symmetry of the dual CFT is then given by the W$_\Gamma$-algebra for the root system $\Gamma$.\footnote{The prescription of extracting W-algebra relations in \cite{Kimura:2015rgi} works beyond the root systems.} Therefore, in the previous example of pure $SU(n)$ super Yang-Mills theory, which in quiver language is an $A_1$ quiver with color label $n$, the dual CFT will have Virasoro (W$_2$) symmetry. In other words, the algebra depends on the rank of the quiver. The dependence on ranks of individual gauge groups appears in Virasoro constraints at level $n$ which need to be included. In the limit $n\to\infty$ these constraint disappears and the one obtains a complete (q)Virasoro algebra.

Therefore, for type $A$ quiver gauge theories in five dimensions, the Kimura-Pestun duality provides a spectral dual (also fiber-base dual or S-dual) version of the BPS/CFT to the standard (q)AGT relation. This can be understood from studying the brane picture for the corresponding gauge theories. We decide to work with 5d theories since K-theoretic Nekrasov functions are better behaved under the above-mentioned duality. Also \cite{Kimura:2015rgi} is naturally formulated in five dimensions. Thus on the CFT side of the correspondence we are dealing with difference or qVOAs.

This paper analyzes the relationship between the standard AGT approach for $SU(n)$ gauge theory adjoint matter in the presence of monodromy defects and the construction of Kimura and Pestun by embedding both theories into a certain gauge origami construction by Nekrasov \cite{Nekrasov:2016ydq} (see \secref{Sec:GaugeOrigami}). We then shall consider both constructions at large-$n$ limit and find that the building blocks of qW algebras and the stable limit of spherical $\mathfrak{gl}_n$ double affine Hecke algebra (DAHA) -- the algebra that acts on the Hilbert space of states of codimension two defects of $\mathcal{N}=1^*$ theory \cite{Koroteev:2018isw} can be identified up to a simple replacement of equivariant parameters in the underlying origami picture. Both algebras are realized in terms of oscillators of doubly-deformed Heisenberg algebra \eqref{eq:qHeisenb}.

The approach to the BPS/CFT correspondence via large-$n$ limit is not new -- the  mathematical proof of the AGT conjecture for A-type theories without fundamental matter
\cite{2012arXiv1202.2756S} uses it abundantly. Later in \cite{Koroteev:2015dja,Koroteev:2016znb} and then in \cite{Koroteev:2018isw} it was used to study large-$n$ behavior of equivariant K-theory of quiver varieties which had an effective description in terms of equivariant K-theory of the Hilbert scheme of points on $\mathbb{C}^2$.
The elliptic Hall algebra, which will appear in \secref{Sec:AlbegraE}, acts on the latter space by means of Nakajima correspondences.

Recently gauge origami and vertex operator algebras was studied systematically from the point of view of  representation theory and algebraic geometry  \cite{Rapcak:2018nsl} (see also \cite{Gaiotto:2017euk,Prochazka:2018tlo} for more discussion of VOAs and defects). The authors have constructed the action of cohomological Hall algebra (COHA) on the moduli space of spiked instantons. In our description COHA reduces to the elliptic Hall algebra $\mathfrak{E}$ and the spiked instantons become folded instantons. In the future one may extend our analysis to a complete origami construction with defects.

\section{Gauge Origami with Defects}\label{Sec:GaugeOrigami}
Consider gauge origami setup introduced by Nekrasov (for more details and notation see \cite{Nekrasov:2016ydq}) for Type IIA string theory on $X\times S^1\times\mathbb{R}$. For our purposes it is enough to take $X=\mathbb{C}^4$ in the presence of Omega background with parameters $\epsilon_1,\epsilon_2,\epsilon_3,\epsilon_4$ such that $\sum_a\epsilon_a=0$. We shall study K-theoretic version of instanton partition function for gauge theories living on D4 branes wrapping $\mathbb{C}^2\times S^1$ for some choice of $\mathbb{C}^2\subset\mathbb{C}^4$.  
Let us denote the number of branes wrapping complex planes $\mathbb{C}_{\epsilon_a}\times \mathbb{C}_{\epsilon_b}$ by $n_{ab}$.

Our starting point is the origami construction with $n_{12}=n$ in the presence of Abelian $\mathbb{Z}_n$ orbifold along two 2-planes as follows
\begin{equation}
\Gamma = \text{diag}(1 \,\,\omega\, \,1\,\, \omega^{-1})\,,
\label{eq:OrbiAbelian}
\end{equation}
where $\omega^n=1$. The 12 plane supports $\mathcal{N}=2^*$ theory with gauge group $U(n)$ in the Omega background with defect along the second plane. Then we need to put some branes in 13 plane, say $n_{13}=M$. This configuration with one common complex line ($\mathbb{C}_{\epsilon_1}$ in this case) is called \textit{folded instantons} \cite{Nekrasov:2016qym}. 

Therefore we get the $\widehat{A}_{n-1}$ necklace quiver with $U(M)$ gauge group at each node supported on $\mathbb{C}_{\epsilon_1}\times\mathbb{C}_{\epsilon_3}$ and $\widehat{A}_{0}$ $U(n)$ theory supported on $\mathbb{C}_{\epsilon_1}\times\mathbb{C}_{\epsilon_2}$ whose adjoint hypermultiplet has mass $\epsilon_3$ in the presence of the monodromy defect along $\mathbb{C}_{\epsilon_1}$. 

\begin{figure}[h!]
\includegraphics[scale=0.5]{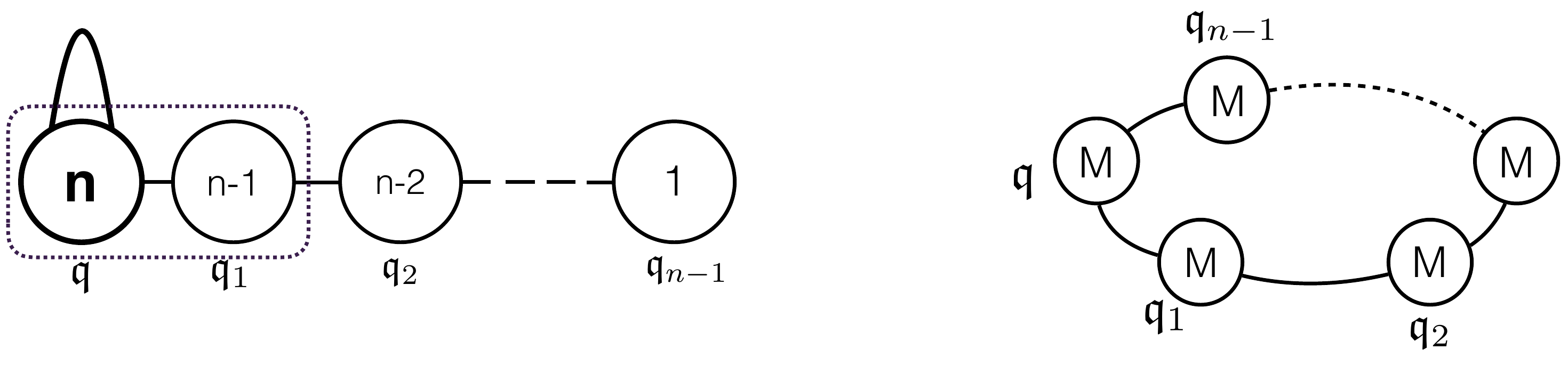}
\caption{Left: 5d $\mathcal{N}=1^*$ (or $\widehat A_0$) theory with three-dimensional full monodromy defect. Right: 5d $\widehat A_{n-1}$ quiver with $U(M)$ gauge groups.}
\label{fig:foldedinstn123}
\end{figure}

Now the orbifolding introduces $n-1$ new parameters $\mathfrak{q}_1, \dots, \mathfrak{q}_{n-1}$.
They will serve as defect parameters (first Chern classes or Fayet-Iliopoulos parameters) on one side and as gauge couplings of the necklace quiver theory on the other, see \figref{fig:foldedinstn123}.

\subsection{Characters and Statistical Weights}
In \cite{Nekrasov:2016ydq} it was defined how to implement orbifolding \eqref{eq:OrbiAbelian} inside the gauge origami calculations. The reader should consult \textit{loc. cit.} for more technical details. First, one computes the $\Gamma$-invariant part of the character on the tangent space to the moduli space of instantons. Assume $A=\{a,b\}$ for $a<b\in \textbf{4}=\{1,2,3,4\}$ and $\omega\in\Gamma^\vee$.
We have the following expressions for the characters
\begin{align}
N_{A,\omega}&= \sum_{\alpha=1}^{n_{A,\omega}} e^{\beta a_{A,\omega,\alpha}},\notag\\
K_{A,\omega}&=\sum_{\widetilde\omega} \sum_{\alpha=1}^{n_{A,\widetilde\omega}}\sum_{(i,j)\in\lambda_{(A,\widetilde\omega,\alpha)}}\delta(\widetilde\omega,\omega-\rho_a(i-1)-\rho_b(j-1))e^{\beta c_{A,\omega,\alpha}}\,,
\end{align}
where  $\rho_a\in\Gamma^\vee$
and the sum is taken over the content of the Young tableau 
\begin{equation}
c_{A,\omega,\alpha}=a_{A,\omega,\alpha}+\epsilon_a(i-1)+\epsilon_b(j-1)\,.
\label{eq:cAdefadd}
\end{equation}
The above is combined into a single character 
\begin{equation}
\widetilde T_{\lambda}^\Gamma = \sum_{A\in\textbf{6},\omega\in\Gamma^\vee} \left(P_{\bar{A},\omega}\left(T^+_{A,-\omega}+\sum_{\omega'\in\Gamma^\vee}N_{A,\omega'}\sum_{B\neq A}K^*_{B,\omega+\omega'}\right)-\sum_{\omega'\in\Gamma^\vee} P_{\textbf{4},\omega'-\omega}\sum_{A<B}K_{A,\omega}K^*_{B,\omega'}\right)\,,
\label{eq:TChar}
\end{equation}
where $T^+_{A,-\omega}$ is the positive half of the orbifolded character
\begin{equation}
T_{A} = N_AK_A^\ast+q_AK_AN_A^\ast-P_AK_AK_A^\ast\,,
\end{equation}
$\bar{A}=\textbf{4}\backslash A$ in the compliment to $\textbf{4}$, also for any $S\subset \textbf{4}$ the character reads
\begin{equation}
P_{S,\omega'-\omega} =  \sum_{J\subset S}\prod_{a\in J}\left(-e^{\beta \varepsilon_a}\right)\delta_{\Gamma^\vee}\left(\omega-\omega'+\sum_{a\in J}\rho_a\right)\,,
\label{eq:P4}
\end{equation}
which is a universal factor and it does not depend on the choice of 2-planes and 
$$
\textbf{6}={\textbf{4}\choose 2}=\{12,13,14,23,24,34\}\,.
$$
In our case with only $n_{12}$ and $n_{13}$ being nontrivial amounts to
\begin{equation}
T_\lambda^\Gamma =\sum_{\omega\in\Gamma^\vee}\left(P_{34,\omega} T^+_{12,-\omega}+P_{24,\omega} T^+_{13,-\omega}\right)+\sum_{\omega,\omega'\in\Gamma^\vee}\left(P_{34,\omega} N_{12,\omega'}+P_{24,\omega} N_{13,\omega'}-P_{\textbf{4},\omega-\omega'}K_{12,\omega}K^*_{13,\omega'} \right)\,.
\label{eq:totalchar}
\end{equation}
The first sum in the above formula has contributions from the $\mathcal{N}=1^*$ theory with the defect and from the necklace theory, whereas the second double sum has mixed terms. The very last terms has instanton contributions from both $n_{12}$ and $n_{13}$ branes.

\subsection{The Partition Function}
Using \eqref{eq:TChar} the full origami partition function reads
\begin{equation}
\mathcal{Z}^\Gamma=\mathcal{Z}^{\text{pert}}\cdot \sum_{\lambda}\left[\prod_{\omega\in\Gamma^\vee}\mathfrak{q}_\omega^{k_\omega}\right]\,\varepsilon\left[- T_{\lambda}^\Gamma\right]\,,
\label{eq:OrigamiPF}
\end{equation}
where $\varepsilon$ translates pure characters into products, $\mathcal{Z}^{\text{pert}}$ is the perturbative contribution and 
\begin{equation}
k_\omega = \sum\limits_{A\in\textbf{6}, \alpha\in n_{A,\omega}}|\lambda_{(A,\widetilde\omega,\alpha)}|\,.
\end{equation}

\subsection{Discrete Symmetry of the Character}
We can notice that there is a symmetry $S_{23}$ of the character \eqref{eq:totalchar} which interchanges complex planes $\mathbb{C}_{\epsilon_2}$ and $\mathbb{C}_{\epsilon_3}$. Moreover, when $n_{12}=n_{23}$ then the character is invariant under this symmetry
\begin{equation}
S_{23}(T_\lambda^\Gamma) = T_\lambda^\Gamma\,.
\label{eq:symmetry23}
\end{equation}
It happens due to the following symmetries of characters \eqref{eq:totalchar} 
\begin{equation}
K^\ast_{13,\omega} = K_{13,-\omega}\,,\qquad K^\ast_{12,\omega} = K_{12,-\omega}
\end{equation}
Finally, $P_{\textbf{4},\omega'-\omega}$ from \eqref{eq:P4} does not depend on the choice of complex lines.

\subsection{Decoupling Limit}
Now let us take the limit $\mathfrak{q}\to 0$. On the left in \figref{fig:foldedinstn123} we have a 3d $\mathcal{N}=2^\ast$ theory on $\mathbb{C}_{\epsilon_1}\times S^1$ whose content is given by a very special Nakajima quiver -- the cotangent bundle to complete flag variety in $\mathbb{C}^n$ with framing given by $GL(n)$ with equivariant parameters $a_{1,1,1},\dots,a_{1,1,n}$ and VEVs of 3d vector multiplets $a_{1,j,n_j},\,  n_j = 1,\dots, n-j,\,j=1,\dots, n-1.$ Here the first index $1\in \textbf{6}$ corresponds to $12$ plane. There $n-1$ adjoint hypermultiplets for each node of the quiver which all have the same mass $\epsilon_3$.

Meanwhile, on the right of \figref{fig:foldedinstn123} the affine quiver turns into finite $A_{n-1}$ quiver since the $U(M)$ gauge group with gauge coupling $\mathfrak{q}$ becomes global. The Coulomb branch parameters of the frozen node are $a_{2,1,i},\, i = 1,\dots, M$, where index $2\in \textbf{6}$ designates the $13$ plane. The 5d gauge couplings for $n-1$ $U(M)$ gauge groups are $a_{2,i,m_i},\, i=1,\dots, n-1,\,m_i = 1,\dots, M$.

\begin{figure}[h!]
\includegraphics[scale=0.45]{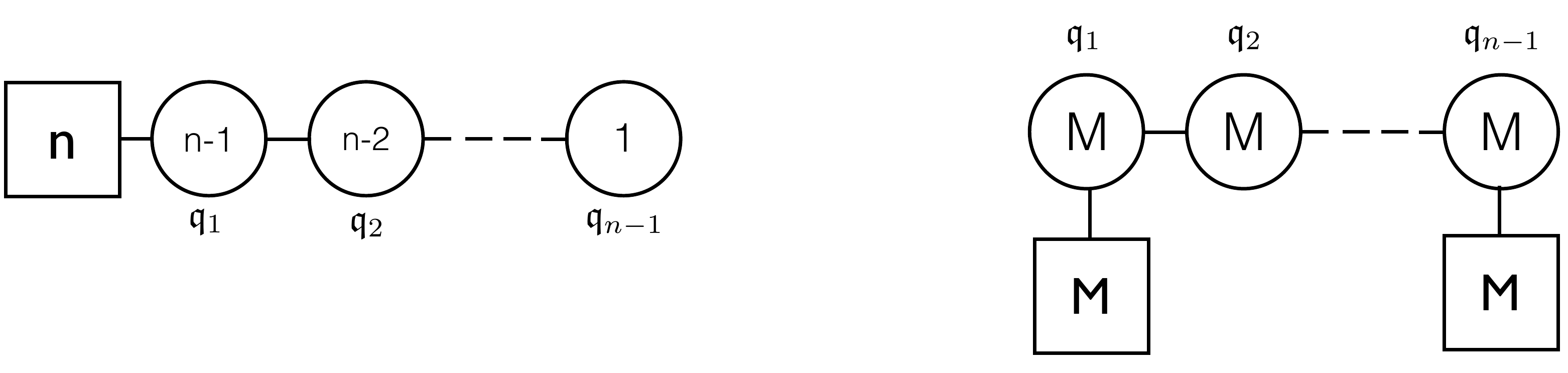}
\caption{After taking the decoupling limit of theories in Fig. \ref{fig:foldedinstn123}. Left: 3d $\mathcal{N}=2^\ast$ quiver gauge theory. Right: 5d framed $A_{n-1}$ quiver with $U(M)$ gauge groups.}
\label{fig:foldedn123decoup}
\end{figure}

In summary the pair of theories from \figref{fig:foldedinstn123} reduces to those depicted in \figref{fig:foldedn123decoup} in $\mathfrak{q}\to 0$ limit.

\subsection{Truncation of Defect Partition Function}
\begin{figure}[!h]
\centerline{
\begin{picture}(200,200)
\put(20,170){{$\lambda^{(1)}$}}
\put(75,115){{$\lambda^{(2)}$}}
\put(185,17){{$\lambda^{(M)}$}}
\multiput(0,155)(5,0){5}{\line(1,0){3}}
\put(0,0){\line(0,1){180}}
\put(0,0){\line(1,0){200}}
\put(0,180){\line(1,0){7}}
\put(7,180){\line(0,-1){5}}
\put(7,175){\line(1,0){3}}
\put(10,175){\line(0,-1){5}}
\put(10,170){\line(1,0){5}}
\put(15,170){\line(0,-1){10}}
\put(15,160){\line(1,0){10}}
\put(25,160){\line(0,-1){5}}
\put(25,155){\line(1,0){30}}
\multiput(55,100)(0,5){4}{\line(0,1){3}}
\multiput(55,100)(5,0){5}{\line(1,0){3}}
\put(55,155){\line(0,-1){35}}
\put(55,120){\line(1,0){5}}
\put(60,120){\line(0,-1){5}}
\put(60,115){\line(1,0){5}}
\put(65,115){\line(0,-1){5}}
\put(65,110){\line(1,0){10}}
\put(75,110){\line(0,-1){5}}
\multiput(165,0)(0,5){4}{\line(0,1){3}}
\put(75,105){\line(1,0){5}}
\put(80,105){\line(0,-1){5}}
\put(80,100){\line(1,0){30}}
\multiput(112,78)(5,-5){5}{\circle*{1}}
\put(110,100){\line(0,-1){20}}
\put(135,55){\line(1,0){30}}
\put(165,55){\line(0,-1){35}}
\put(165,20){\line(1,0){5}}
\put(170,20){\line(0,-1){5}}
\put(170,15){\line(1,0){10}}
\put(180,15){\line(0,-1){5}}
\put(180,10){\line(1,0){5}}
\put(185,10){\line(0,-1){5}}
\put(185,5){\line(1,0){15}}
\put(200,5){\line(0,-1){5}}
\end{picture}}
\caption{The asymptotic partition $\Lambda$.}
\label{Fig:AsymptPart}
\end{figure}
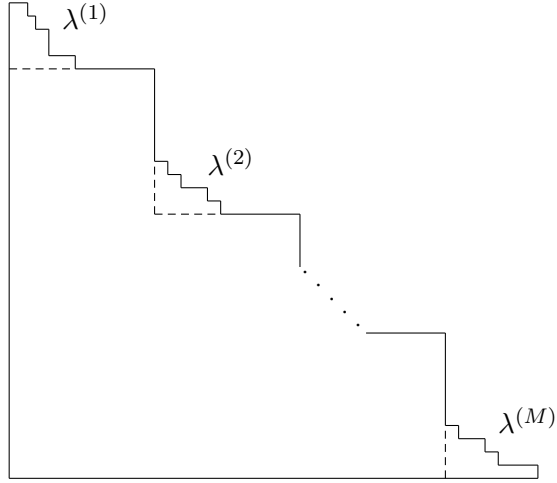

Consider the following locus in the space of variables $a_{1,1,j},\, j= 1,\dots n$
\begin{equation}
a^{(\ell)}_{1,1,i} =\mathrm{a}_\ell q_1^{\lambda^{(\ell)}_i}q_2^{i-n}\,,\qquad i = 1,\dots,n\,,\quad \ell = 1,\dots, M\,,
\label{eq:HiggsingCodim2A}
\end{equation}
where $\mathrm{a}_\ell$ are arbitrary $\mathbb{C}^\ast$-valued parameters. Here we switched from additive notations of \eqref{eq:cAdefadd} to multiplicative conventions: $q_a=e^{\beta\epsilon_a}$ and $\mathrm{a}_\ell=e^{\beta a_\ell}$.
Then, as it was shown in \cite{Koroteev:2018isw}, the defect partition function (or equivariant K-theory vertex function) of the theory on left in \figref{fig:foldedn123decoup} truncates to a generalized Macdonald polynomial
\begin{equation}
\mathcal{Z}_{\text{defect}}\big\vert_{\eqref{eq:HiggsingCodim2A}} = P_{\vec{\lambda}} (\mathfrak{q}_1,\dots, \mathfrak{q}_{n-1}; q_1, q_3)\,,
\label{eq:DefectPF}
\end{equation}
for M-tuple partition $\lambda = (\lambda^{(1)},\dots,\lambda^{(M)})$ (see sec. 4.2 of \cite{Koroteev:2018isw}). 
These $M$ partitions can be blended into the so-called asymptotic partition \figref{Fig:AsymptPart}
so that constraints \eqref{eq:HiggsingCodim2A} can be written as 
\begin{equation}
a_{1,1,i} =\mathrm{a}_i q_1^{\Lambda_i}q_2^{i-n}\,,\qquad i = 1,\dots,n\,.
\label{eq:HiggsingCodim2}
\end{equation}
The total number of columns of $\Lambda$ is equal to $n$. We assume that the number of columns in each partition $\lambda^{(\ell)}$ is parametrically smaller than $n/M$.

These asymptotic partitions provide bases in the equivariant K-theory of the moduli space of $M$ $U(n)$ instantons $\mathcal{M}_{M,n}$.

\subsection{Higgsing and Truncation of $\mathcal{Z}$-state.}
Let us impose the following conditions on Coulomb VEVs of the 5d gauge theory in \figref{fig:foldedn123decoup} 
\begin{equation}
\frac{a_{2,i+1,m_i}}{a_{2,i,m_i}} = q_1^{\mu^{(i)}_{m_i+1}-\mu^{(i)}_{m_i}}q_3\,,\qquad i = 1,\dots,n-1\,,
\label{eq:Higgsing}
\end{equation}
where values $i=1$ and $i=n+1$ correspond to fundamental matter fields with VEVs $a_{2,1,i},\, i = 1,\dots, M$.
Equivalently the above formula can be represented as
\begin{equation}
a_{2,i,m_i} = \mathrm{a}_{2} q_1^{\mu^{(i)}_{m_i}}q_3^{i-\lfloor \frac{n}{M} \rfloor}\,,\qquad i = 1,\dots,n-1\,,
\label{eq:Higgsing}
\end{equation}
where $\mathrm{a}_{2}$ is an overall scale and by construction $\mu^{(1)}_{m_i}\leq \mu^{(2)}_{m_i}\leq\dots\leq \mu^{(n)}_{m_i}$. 

As it has been discussed in a number of publications\footnote{Some of these papers work with 4d theories and some work with the Nekrasov-Shatashvili limit $q_2\to 1$.} \cite{Dorey:2011pa,Chen:2011sj,Bulycheva:2012ct,Chen:2013jtk,Aganagic:2014oia,Aganagic:2013tta} condition \eqref{eq:Higgsing} defines (baryonic) locus of the Higgs brach of the quiver gauge theory in question. Indeed, the instanton partition function $\mathcal{Z}$ of the 5d theory truncates into a 3d vortex partition function of the defect theory. Note that it's a different kind of defect than the one we just talked about. In M-theory approach of \cite{Alday:2010vg} this is a \textit{codimension four} defect vs. the \textit{codimension two} defect in the $\mathcal{N}=1^*$ theory.

\subsection{Matching}
In the Higgsing procedure which we have described above the $A_{n-1}$ quiver yields an $M\times n$ matrix of integers $\mu^{(i)}_{m_j}$, where $i=1,\dots, n$ and $j = 1,\dots, M$. There are two natural ways to combine those numbers into partitions -- combining matrix elements in rows or in columns into Young diagrams. If we form Young diagrams by combining the matrix elements in rows we shall get an $n$-tuple of tableaux: $\mu_{m_i}=\left\{\mu^{(1)}_{m_i},\dots,\mu^{(n)}_{m_i}\right\}$.
Then we can construct an $M$-tuple of such partitions $\mu=\left\{\mu_{m_1},\dots,\mu_{m_M}\right\}$.

We can see that \eqref{eq:Higgsing} matches with \eqref{eq:HiggsingCodim2} upon identifying $q_2$ with $q_3$ and $\lambda = \left\{\lambda^{(1)},\dots,\lambda^{(M)}\right\}$ with $\mu$ provided that $\mathrm{a}_\ell=\mathrm{a}_2 q_3^{-\lfloor \frac{n}{M} \rfloor}$ for all $\ell$.
Note that not all $\lambda^{(i)}_{j}$ (or $\mu^i_{m_j}$) must be nontrivial.

To summarize, the proposed duality works as follows. From the folded instantons construction with branes along $n=n_{12}$ and $M=n_{13}$ directions we introduce a defect along complex line $\mathbb{C}_{\epsilon_1}$ by adding a $\mathbb{Z}_n$ orbifold $\Gamma$ along directions $2$ and $4$. Due to the symmetry between $2$ and $3$ directions we conclude that in the decoupling limit $\mathfrak{q}\to 0$ and upon imposing `quantization conditions' \eqref{eq:HiggsingCodim2} (equivalently \eqref{eq:Higgsing}) the origami partition function \eqref{eq:OrigamiPF} turns into a generalized Macdonald polynomial $P_{\vec{\lambda}}$.

\subsection{Fourier-Mukai Transform}\label{Sec:FourierMuk}
The brane realization of the 5d $A_{n-1}$ quiver gauge theory with $U(M)$ gauge groups involves intersecting NS5 and D5 branes. If we rotate this picture by 90 degrees, thereby interchanging NS5 and D5 branes, then the gauge theory limit of the rotated brane picture will be the $A_{M-1}$ quiver theory with $U(n)$ gauge (and flavor) vertices. 
On the Higgs branch locus \eqref{eq:Higgsing} we need to transpose matrix $\left[\mu^{(i)}_{m_j}\right]$.
The transposed partitions therefore describe the data of \textit{bona fide} codimension four defect in the $A_{M-1}$ quiver gauge theory with framing. 

To summarize, the partition function of the maximal monodromy defect in $U(n)$ 5d $\mathcal{N}=1^*$ theory on locus \eqref{eq:HiggsingCodim2} can be identified with the codimension four defect of the 5d $A_{M-1}$ quiver gauge theory. 

Notably we have demonstrated that a codimension-two defect is related to a codimension-four defect by a brane rotation which is a manifestation of the Fourier-Mukai transform in string theory (cf. \cite{Frenkel_2016}).

\section{Quiver qW-algebra in the Large-$n$ Limit}
In this final section we shall address some aspects of representation theory which naturally arise from studying the gauge origami construction from \secref{Sec:GaugeOrigami}. In particular, we can ask what happens if $n_{12}=n$ becomes large. 

The large-$n$ limit of the left part of \figref{fig:foldedn123decoup} was studied in details in \cite{Koroteev:2018isw}. The defect partition function \eqref{eq:DefectPF} can be written in the free boson formalism as
\begin{equation}
\mathcal{Z} = \widehat{\mathcal{Z}}|0\rangle
\end{equation}
and operators built of creation and annihilation operators acting on the Fock vacuum. This operators are in one-to-one correspondence with power-symmetric combinations of $\mathfrak{q}_1,\dots, \mathfrak{q}_{n-1}$-variables and
form a two-parameter family deformation of the Heisenberg algebra
\begin{equation}
[a_n,a_m]=m\frac{1-q_1^{m}}{1-q_3^{m}}\delta_{m,-n}\,.
\label{eq:qHeisenb}
\end{equation}
The same exact commutation relations are obeyed by the creation and anninilation operators used in \cite{Kimura:2015rgi} to construct their $\mathcal{Z}$-states. By now the similarity between constructions by us and Kimura and Pestun for type-A quivers should not be surprising. Here we would like to comment more on the algebraic aspects of this connection.

\subsection{Algebra $\mathfrak{E}$}\label{Sec:AlbegraE}
The following algebras which depend on two parameters $q_1,q_2\in\mathbb{C}^\times$ are isomorphic to each other
\begin{equation}
U_{q_1,q_2}\left(\widehat{\widehat{\mathfrak{gl}_1}}\right) \simeq \mathscr{E}_{q_1,q_2} \simeq \mathfrak{gl}_\infty \text{DAHA}_{q_1,q_2}^S \simeq \text{DIM}_{q_1,q_2} \simeq D(\mathscr{A}_{\text{shuffle}})\,,
\label{eq:allalgebras}
\end{equation}
which are respectively: quantum toroidal $\mathfrak{gl}_1$ algebra, elliptic spherical Hall algebra, large$-n$ stable limit of spherical double affine Hecke algebra of $\mathfrak{gl}_n$, Ding-Iohara-Miki algebra, and Drinfeld double of shuffle algebra. For simplicity we shall call algebras from \eqref{eq:allalgebras} $\mathfrak{E}$ which plays the major role in this development. The reader can consult \cite{Negut:2012aa} for side-by-side comparison of the algebras from \eqref{eq:allalgebras}.

Algebra $\mathfrak{E}$ is a Hopf algebra with central element $\gamma$ (such that its coproduct $\Delta (\gamma) = \gamma\otimes\gamma$). According to \cite{cite:FeiginShiraishi} $\mathfrak{E}$ has representation $\rho$ on the Fock space of Heisenberg algebra \eqref{eq:qHeisenb} by specifying $\rho(\gamma)=(q_1q_3)^{-\frac{1}{2}}$. This is the so-called level one representation. Representations of $\mathfrak{E}$ for other levels such that $\rho(\gamma)=(q_1q_3)^{-\frac{M}{2}}$ lead to deformed qW$_{M}$ algebras. 

\subsection{Spherical DAHA}
It was shown in \cite{2012arXiv1202.2756S} that there exist the following surjective algebra homomorphism from $\mathfrak{E}$ to spherical $\mathfrak{gl}_n$ DAHA  which we shall call $\mathfrak{A}_n$ for short.
\begin{equation}
\mathfrak{E} \longrightarrow \mathfrak{A}_n\,.
\end{equation}
Algebra $\mathfrak{A}_n$ can be thought of as a geometric quantization of the moduli space of flat $GL(n;\mathbb{C})$ connections on a punctured torus (the $n$-particle Calogero-Moser space) in complex structure $J$ \cite{A.Oblomkov:}, with a simple monodromy around the punction with eigenvalue $\beta^{-1}\log q_3$. 

We showed in \cite{Koroteev:2018isw} that $\mathfrak{A}_n$ acts on the equivariant K-theory of a quiver variety which corresponds to the 3d gauge theory \eqref{eq:DefectPF}. In the $n\to\infty$ limit this vector space becomes $\mathbb{C}_{q_1}^\times\times\mathbb{C}_{q_3}^\times$-equivariant K-theory of the moduli space of $U(1)$ instantons $K_{q_1,q_3}\left(\mathcal{M}^{\text{inst}}_{k,U(1)}\right)$ on which algebra $\mathfrak{E}$ acts by correspondences.

\subsection{Quiver qW Algebra}\label{Sec:QuiverWRev}
Pestun and Kimura \cite{Kimura:2015rgi,Kimura:2016dys} have a construction of qW-algebras from the free-boson formulation as a commutant of screening charges. We provide a quick review of their construction here.
One starts with a balanced finite or affine $\Gamma=$ADE-type quiver gauge theory in 4,5 or 6 dimensions with node vector $\textbf{n}$. The qW-algebra for root system $\Gamma$ is then an $\epsilon_2$-deformation of the ring of commuting Hamiltonians of the corresponding quantum integrable system studied in \cite{Nekrasov:2013aa}. 
The building block of a qW-algebra is the universal sheaf defined as
\begin{equation}
Y = N - P K\,,
\end{equation}
where $P = (1 - q_1)(1 - q_3)$, and $N$ and $K$ are the sheaves on the moduli space. Then, in terms of `$x$-variable' it is rewritten as
\begin{align}
Y &= (1 - q_1) \sum_{x \in \mathcal{X}} x\,,\notag\\
\mathcal{X} &= \{ x_{\alpha,k} \}\,, \qquad \alpha = 1,\dots, n, \quad k = 1,\dots, \infty\,,
\end{align}
where 
\begin{equation}
x_{\alpha,k} = q_3^{\lambda_{\alpha,k}} q_1^{k-1} a_\alpha\,,
\end{equation}
where $\lambda_{\alpha,k}$ are heights of Young diagram $\lambda_{\alpha}$.
This $x$-variable can be identified with Bethe roots in the Nekrasov-Shatashvili limit $q_3\to 1$. Applying the Adams operation to the $Y$-bundle, we have
\begin{equation}
Y^{[p]} = (1 - q_1^p) \sum_{x \in \mathcal{X}} x^p\,,
\end{equation}
In the 4d $\mathcal{N} = 2$ theory language, this yields a chiral ring operator $Y^{[p]} \sim \text{Tr}\, \Phi^p$.
Thus $Y^{[p]}$ is a $p$-th power-symmetric function of $x$-variables which truncates to a polynomial for finite ranks of gauge groups of the quiver, where $\sum x^p \leftrightarrow a_{-p}$. The above chiral observables are used to construct an extended partition function of quiver theory for root system $\Gamma$ by including the following term in the holomorphic equivariant Euler characteristic
\begin{equation}
\exp\left(\sum\limits_{\alpha=1}^{n_i} t_{i,\alpha} Y_i^{[p]}\right)\,,
\end{equation}
for each vertex of rank $n_i$ (later these ranks should be sent to infinity). Here $t_{i,\alpha}$ are called \textit{higher times} and are in one-to-one correspondence with defect fugacities as we discussed in previous section. For the main example of $A_1$ quiver with a single $U(n)$ node higher times $t_1,\dots, t_{n}$ under our correspondence are directly related\footnote{After projecting out the `center of mass' coordinate} to FI couplings of the 3d defect $\mathfrak{q}_1,\dots, \mathfrak{q}_{n-1}$. The correspondence between defect couplings and chemical potentials for higher Casimir invariants was discussed in the literature before \cite{Alday:2010vg,Frenkel_2016}, here we observe this relationship between two theories depicted in \figref{fig:foldedn123decoup}.

If quiver $\Gamma$ is unbalanced, i.e. finite A-type quiver without fundamental matter, then one can always make it balanced by shifting higher times $t_{i,\alpha}$ by an amount proportional to the masses.

The expansion coefficients of the qq-character for the quiver gauge theory generate the qW-algebra
\begin{equation}
T_i(x)|\CZ\rangle = (x^{n_i}+T_{1,i} x^{n_i-1}+\dots+T_{n_i,i})|\CZ\rangle\,,
\end{equation}
for each vertex of the quiver labelled by $i$ and where instanton partition function $|\mathcal{Z}\rangle$ is represented as a Fock vector in the Hilbert space of states of the Heisenberg algebra \eqref{eq:qHeisenb}. The entireness of the qq-character in terms of the free boson realization can be rephrased as the commutation relation with the screening charges $S_j$ for each node of the quiver. 
\begin{equation}
[T_i(x), S_j(x')] = 0\,.
\end{equation}

Equivalently, one can think of qW algebra for quiver $\Gamma$ whose gauge groups have ranks $n_i$ as geometric quantization of the moduli space $\mathcal{M}^{\text{mon}}_{{\textbf{n}},G_\Gamma}$ of periodic $G_\Gamma$-monopoles on $\mathbb{C}^\times \times S^1$ of charge $\sum_i n_i \alpha_i^\vee$ in complex structure with twistor parameter $\zeta = \frac{\log q_3}{\log q_1}$ \cite{Frassek:2018try,Elliott:2018yqm}. An important remark is that for finite ranks of gauge groups $n_i$ we get all relations in the qW-algebra modulo Virasoro constraints
\begin{equation}
\sum c_{i}T_{-n,i} |\mathcal{Z}\rangle =0\,,\qquad n>n_i\,.
\label{eq:VirConstr}
\end{equation}

In our main example of $A_1$ quiver there is a single gauge group $U(n)$. Therefore the quantization of the moduli space of periodic $SU(2)$ monopoles of charge $n$ yields qW-algebra for $A_1$, or qVirasoro algebra, modulo constraint $T_{-l}=0$ for $l>n$. The Kimura-Pestun qVirasoro algebra acts on the Hilbert space spanned by $|\mathcal{Z}\rangle$-vectors of the $A_1$ quiver theory.

Once the limit $n\to\infty$ is assumed the quantization procedure yields the full qW-algebra.
One can regard the result of \cite{Kimura:2015rgi} as a derivation of a version of the BPS/CFT correspondence, which is different from the standard AGT duality which involves qToda CFT for the root system provided by the quiver diagram of the corresponding gauge theory. We have demonstrated that Kimura-Pestun's and the standard AGT dictionaries are related to each other via a Fourier transform. 

From this perspective the BPS/CFT correspondence can be regarded as a large-$n$ duality for quiver theories of type-A.

\subsection{Comparison}
It is known that Macdonald polynomials \eqref{eq:DefectPF} appear as null vectors in qW algebras \cite{Awata:1995zk,Shiraishi_1996}. 
This also implies that Macdonald (or trigonometric Ruijsenaars-Schneider) operators can be constructed from qW-algebra generators which can be done by using free boson representation of Macdonald operators. Thus for a given level $k$ there will be a relationship of the form \eqref{eq:VirConstr}. Therefore, in terms of Kimura and Pestun's construction, these conditions can be interpreted as Virasoro constraints.

We can also see the similarity of the moduli spaces which arise in deformation quantizations described above. The qW-algebra arises from the space of periodic $SU(2)$ monopoles $\mathcal{M}^{\text{mon}}_{n,SU(2)}$, whereas $\mathfrak{E}$ acts on the moduli space of $U(1)$ instantons $\mathcal{M}^{\text{inst}}_{k,U(1)}$. These two spaces are related via a Fourier transform which replicates the discussion from \secref{Sec:FourierMuk}.

\subsection*{Acknowledgements}
I would like to thank Vasily Pestun, Taro Kimura and Nikita Nekrasov for stimulating discussions.
I gratefully acknowledge support from the Simons Center for Geometry and Physics, Stony Brook University at which some or all of the research for this paper was performed.
I also acknowledge support of IH\'ES and funding from the European
Research Council (ERC) under the European Union's Horizon 2020
research and innovation program (QUASIFT grant agreement 677368). 
The results of this paper were reported at various seminars \cite{Sanyaworkshop,AspenConference2019}.

\bibliography{cpn1}
\end{document}